\documentclass{hotnets2020}

\pdfoutput=1
\usepackage{times}  
\usepackage{hyperref}
\usepackage{enumitem,kantlipsum}
\usepackage{microtype}
\usepackage{balance}

\usepackage{xcolor}

\hypersetup{pdfstartview=FitH,pdfpagelayout=SinglePage}

\usepackage{enumitem}
\newcommand{\squishlist}{\begin{itemize}[itemsep=1pt,parsep=2pt,topsep=3pt,partopsep=0pt,leftmargin=0em, itemindent=1em,labelwidth=1em,labelsep=0.5em]}
\newcommand{\squishend}{\end{itemize}}
\newcommand{\squishenum}{\begin{enumerate}[itemsep=1pt,parsep=2pt,topsep=3pt,partopsep=0pt,leftmargin=0em,listparindent=1.5em,labelwidth=1em,labelsep=0.5em]}
\newcommand{\squishsubenum}{\begin{enumerate}[itemsep=1pt,parsep=2pt,topsep=0pt,partopsep=0pt,leftmargin=0em,listparindent=1.5em,labelwidth=1em,labelsep=0.5em]}
\newcommand{\squishenumend}{\end{enumerate}}

\begin{document}



\title{Open Research Problems in Backscatter Networking}

\title{Advances and Open  Problems in Backscatter Networking}

\author{Vamsi Talla$^{2}$, Joshua Smith$^{1,2}$ and Shyamnath Gollakota$^{1,2}$\\
\\
\normalsize{$^{1}$University of Washington, $^{2}$Jeeva Wireless}\\

}

\maketitle

{\bf Abstract ---} Despite significant research in backscatter communication over the past decade, key technical open problems  remain under-explored. Here, we first systematically lay out the design space for backscatter networking and identify applications that make backscatter an attractive communication primitive. We then identify 10  research problems that remain to be solved in backscatter networking. These open problems span across the network stack to include circuits, embedded systems, physical layer, MAC and network protocols as well as applications. We believe that addressing these problems can help deliver on backscatter's promise of low-power ubiquitous connectivity.\footnote{To appear in ACM GetMobile.}

\section{\normalsize\bf{Introduction}}

The last decade have seen significant advances in backscatter networking research.  Our community has demonstrated backscatter systems that inter-operate with Wi-Fi~\cite{nsdi16,wifibackscatter}, Bluetooth, Zigbee~\cite{interscatter}, FM~\cite{fmbackscatter}, LoRa~\cite{lorabackscatter} and  TV~\cite{abc}. Multiple novel sensing systems like battery-free  phones~\cite{batteryFreePhone}, cameras~\cite{nsdi2018camera} and plastic objects that can talk Wi-Fi~\cite{printedwifi} have also been demonstrated that have captured the imagination of the broader public, industry and academia.

However, given the pace of innovation in this space, a systematic case for where backscatter can be beneficial has not been explicitly made. Given the historic association of backscatter with commercial battery-free RFID systems that harvest power from radio frequency (RF) signals, backscatter has been often conflated with RF power harvesting and battery-free computing. More importantly, the network architectures used and assumptions made across various papers differ significantly, raising a number of unanswered research questions.

In this article, we first make a case for decoupling the academic conversation about backscatter communication from RF power harvesting. We explore the design space of backscatter and make the case for why it is useful as a low-power communication mechanism for Internet of things (IoT) devices with small batteries or non-RF harvesting modalities.  We then  identify three classes of open research problems that remain to be solved.  We list challenges that we see with existing backscatter systems and the various gaps in their capability that as practitioners we believe hinder adoption. We identify future opportunities and directions for backscatter research that explore trends in the industry  as well as innovative opportunities including Terahertz backscatter communication, high-resolution video streaming using backscatter as well as acoustic backscatter for  everyday  IoT devices to  communicate over surfaces or in air.  we outline also various exciting applications like programmable smart dust and connectivity for computational materials that can  deliver on backscatter's  promise of low-power ubiquitous connectivity for the next billion devices.

\begin{figure}[t!]
{\footnotesize
	\begin{tabular}{c}
	\hspace{-0.2in}
	 \includegraphics[width=\columnwidth]{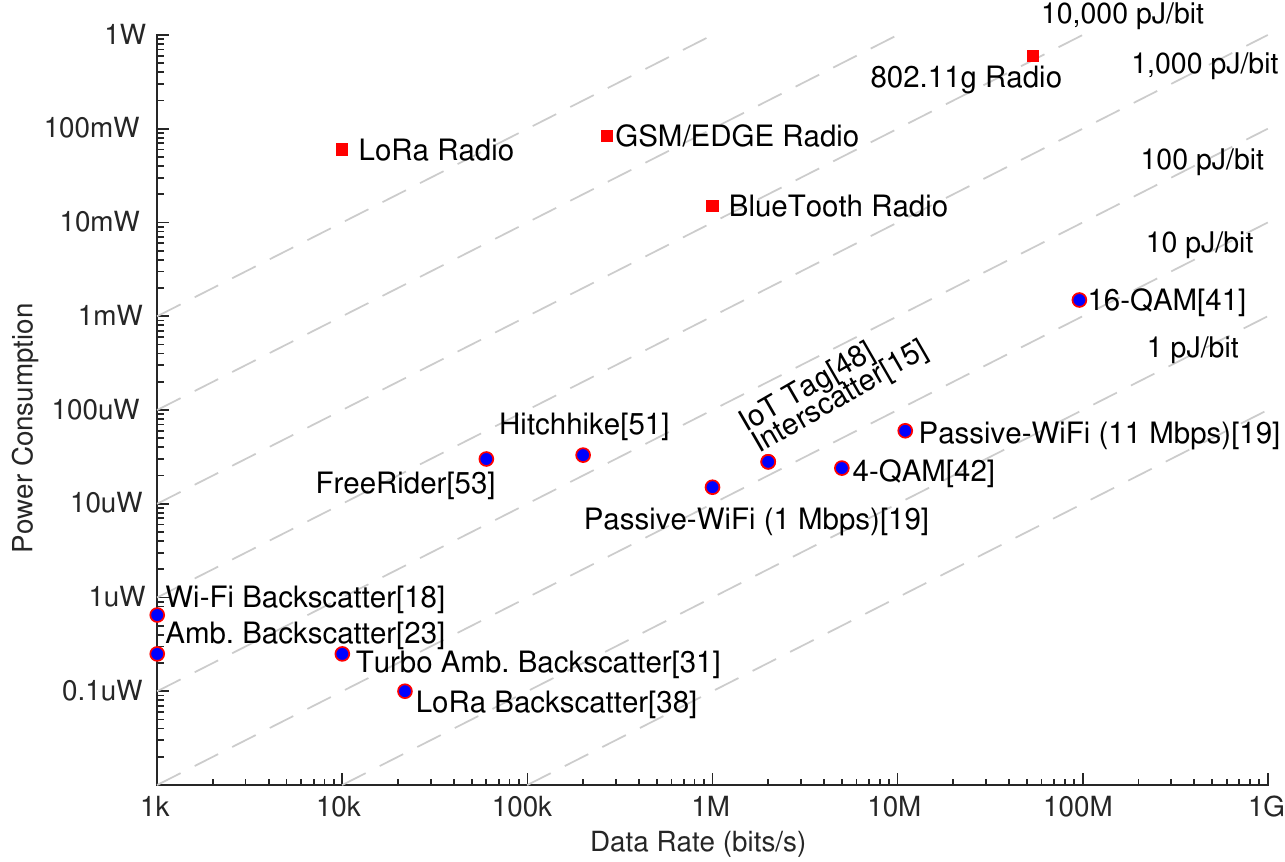}

	\end{tabular}
}
	\vskip -0.2in
	\caption{\textbf{Energy efficiency.}  \textnormal{We plot power consumption of various radio and backscatter communication technologies as a function of data rates. The dotted lines  represent constant energy efficiency.}}
	\label{fig:energy_efficiency}
	\vskip -0.23in
\end{figure}

\section{\normalsize\bf{Decoupling Backscatter \& RF Power}}
\label{sec:case}
Backscatter communication has often been coupled with RF power harvesting and battery-free devices given its long history with passive RFID tags. However, this greatly limits the scope and design space for backscatter communication. RF harvesters have limited sensitivity of -18 to -33~dBm when they need to power the device from cold start. In contrast,  backscattered data packets can be decoded by LoRa receivers down to -134~dBm at distances of up to 2.8~km~\cite{lorabackscatter}. If we can free backscatter communication from the unreasonable constraints of RF energy harvesting and power delivery, it has the potential to replace radios in IoT devices with small batteries or non-RF harvesting modalities. 

First, we outline three design parameters which would help guide the choice between radios and backscatter communication.

\vskip0.05in \noindent{\it i) Energy efficiency.} The key value proposition of backscatter is ultra-low power consumption. However, to make a fair comparison between any two communication technologies, we also need to consider data rates. For example, transmissions at higher power but for shorter duration of the time can be more energy efficient than transmitting at lower power for longer periods. Hence, energy efficiency, i.e., energy per bit is a better metric. In Fig.~\ref{fig:energy_efficiency}, we plot the active power consumption for radios and backscatter as a function of data rates. The dotted lines at an angle represent constant energy efficiency. It can be seen that backscatter systems are at least 1000 times more energy efficient than corresponding radios. 

\begin{figure}[t]

{\footnotesize
	\begin{tabular}{cc}
	\hspace{-0.1in}
	\epsfig{file=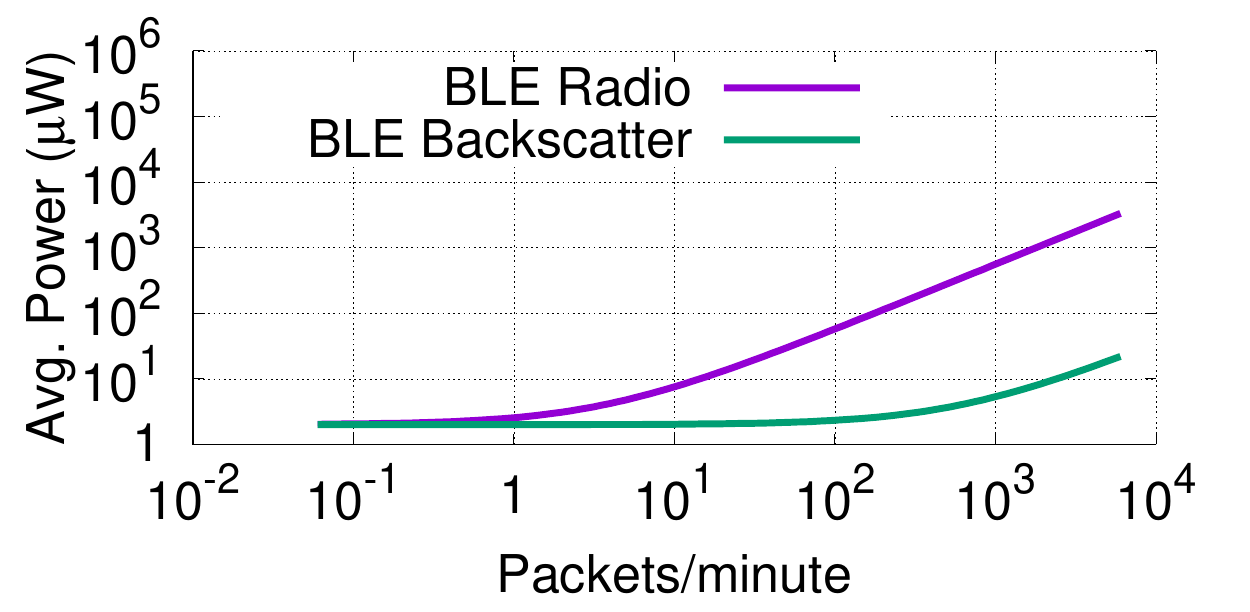, width= 0.5\columnwidth} & 
	\hspace{-0.2in}
	\epsfig{file=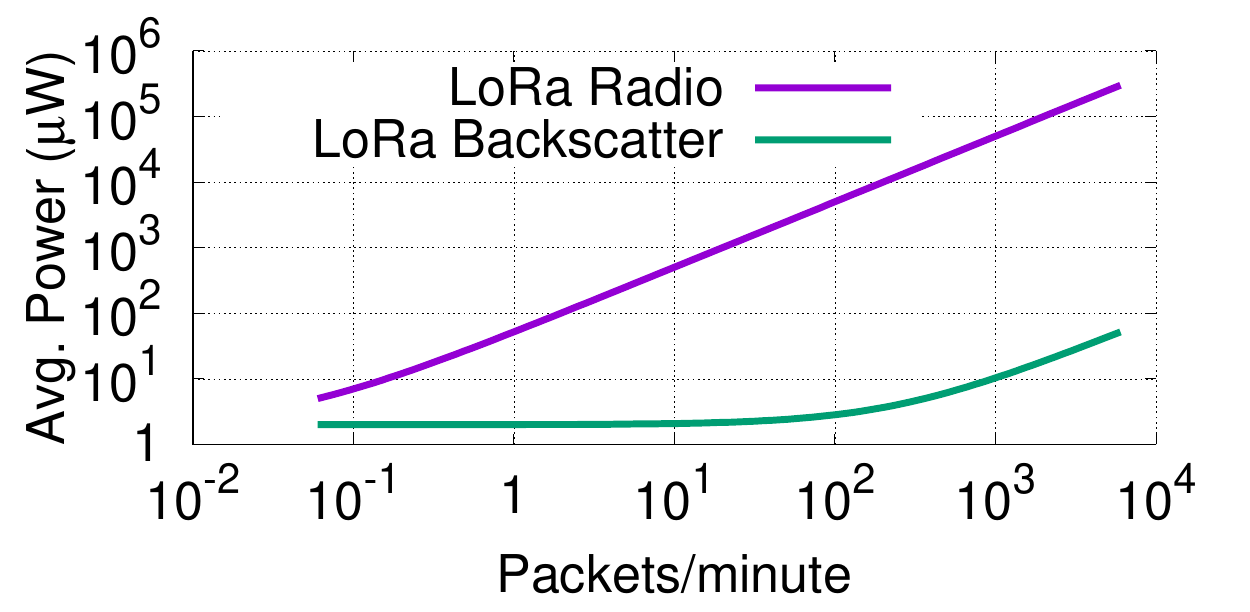, width= 0.5\columnwidth}\\
	(a) BLE protocol & (b) LoRa protocol
	\end{tabular}
}
	\vskip -0.2in
	\caption{\textbf{Power consumption versus duty cycling.} \textnormal{We compare BLE and LoRa radios and their backscatter counterparts.}}
	\label{fig:duty_cycle} 	
	\vskip -0.2in
\end{figure}

\vskip0.05in \noindent{\it ii) Power consumption versus duty cycling.} Energy efficiency only accounts for the active power consumption of communication. However, typical sensor systems invariably transmit data for some time and spend majority of time doing other operations or in sleep mode. To account for this behavior, we need to study the impact of duty cycling on overall power consumption of the system. In order to simplify the analysis of backscatter and radios, we consider a system whose only task is to operate in sleep mode and periodically wake up to transmit data. In Fig.~\ref{fig:duty_cycle}, we plot the average power consumption as a function of packets transmitted per minute for LoRa and Bluetooth radios (at 0~dBm) and backscatter counterparts. It can be seen that at low packet transmission rates the power consumption of radios and backscatter is comparable as the power is dominated by the quiescent current (2~$\mu$A) whereas at higher transmit rate, the superior energy efficiency of backscatter outperforms by multiple orders of magnitude. Additionally, state of the art Bluetooth radios are more energy efficient than LoRa radios due to their low active power (15 vs 60~mW) and shorter packet length (100-400~$\mu$s vs 10-100~ms) which explains the fact that Bluetooth radios are comparable to a backscatter design at update rates of 1 packet per second whereas backscatter is more energy efficient than a LoRa radio even at 1 packet every 10 minutes. That said, backscatter can significantly outperform Bluetooth for high duty cycle application. For example, an interactive application such as streaming accelerometer data at a rate of 10 Hz with BLE beacons will consume 335~$\mu$W, while backscatter can support the same application with just 4~$\mu$W. This analysis only considers packet transmission and a more holistic analysis should consider the impact of receiver, packet re-transmissions, computation and sensing on the overall power consumption. 

\begin{figure}[t]

{\footnotesize
	\begin{tabular}{cc}
	\centering
	\hspace{-0.1in}
	\epsfig{file=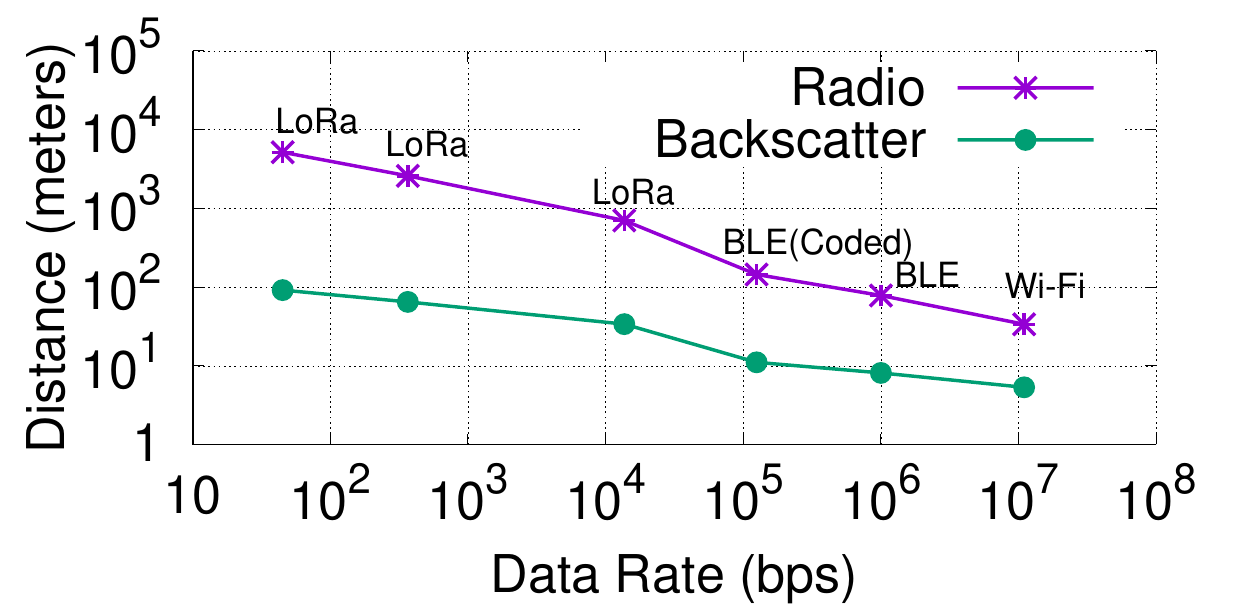, width= 0.5\columnwidth} & 
	\hspace{-0.2in}
	\epsfig{file=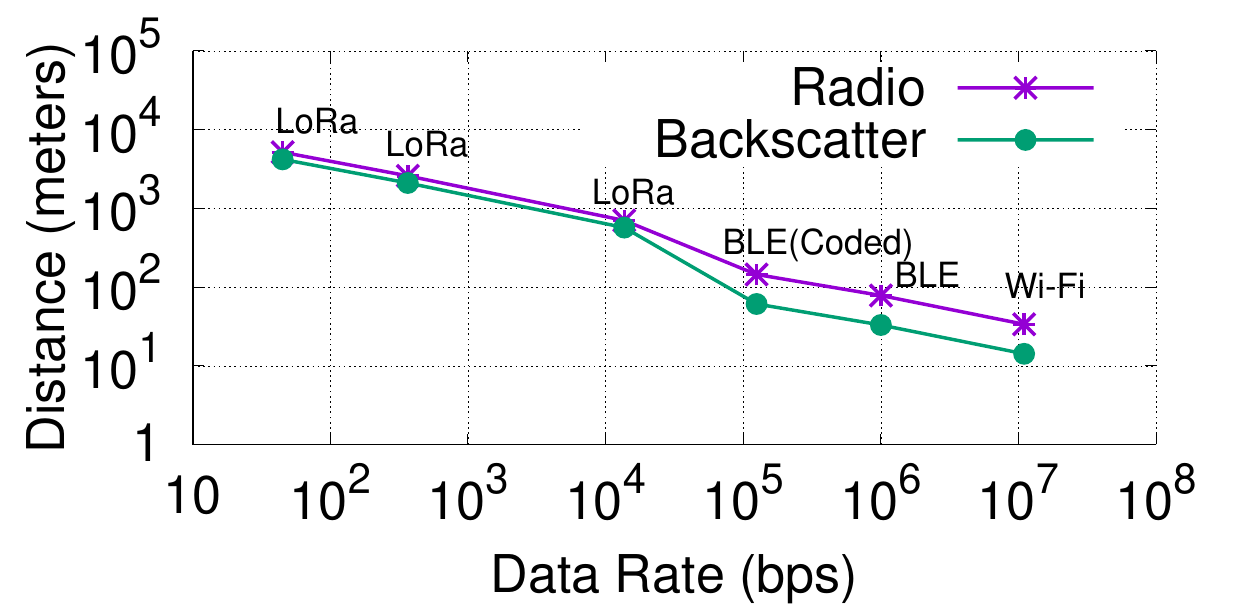, width= 0.5\columnwidth} \\
	\includegraphics[page = 1, width= 0.35\columnwidth]{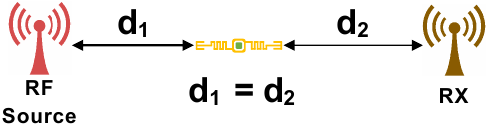} & 
		//\hspace{-0.2in}
    \includegraphics[page = 2, width= 0.35\columnwidth]{./figures/fig_deployment-crop.pdf} \\
	(a) Device in the middle & (b) Device close to RF source
	\end{tabular}
}
	\vskip -0.2in
	\caption{\textbf{Operating distances.} \textnormal{We compare radios with two backscatter deployments.}}
	\label{fig:distance_data_rate} 	
	\vskip -0.2in
\end{figure}

\vskip0.05in \noindent{\it iii) Operating distances.} Finally, although backscatter is at least three orders of magnitude lower power than radios, there is a trade-off in operating distances due to the dual path loss propagation in backscatter systems. To demonstrate this trade-off, we plot the maximum operating distances of radios and backscatter versions of Wi-Fi, Bluetooth and LoRa protocols as a function of data rate in Fig.~\ref{fig:distance_data_rate}. The plots are based on Friis path loss model for a propagation exponent of 3, which is a conservative setting with significant multi-path and attenuation due to walls indoors. The radios transmit at 0~dBm to minimize power consumption whereas the carrier source in backscatter is configured to transmit at 30~dBm since carrier source is typically an access point plugged in. 


\vskip 0.05in\noindent{\bf Analysing application scenarios.}
The choice between radios and backscatter communication for an application is a function of operating distance, data rate requirement and the power budget. There is currently  no one size fits all solution~\cite{braidio}. {For example, if one wants to achieve 1 Mbps at 20--40 m, Bluetooth radio is likely the only option (barring future innovations in backscatter). But backscatter is a much better fit for an application that requires 100 kbps at 5--10 m, since it provides orders of magnitude better energy efficiency.} We consider three different power scenarios and outline a methodology to help navigate this design decision. 

\vskip0.05in \noindent{\it i) Small batteries.} A system powered by small batteries has a limited power budget and invariably if a radio is used, it will dominate the power budget. Consider a temperature sensor powered by 10~mm CR1025 coin cell and transmitting a packet every minute to a receiver placed at a distance of 100 meters. We can see that LoRa would meet the bit rate and operating distance requirement. For a packet transmitted every minute, a LoRa radio would consume an average power of 0.5~mW whereas backscatter consumes less than 2.1~$\mu$W. The CR1220 coin cell has a nominal capacity of 30~mAhr which translates to operation for 7.5~days for a LoRa radio versus the 5 years lifetime for a backscatter solution. This is just a first order analysis as the capacity of the battery is also a function of peak current draw and discharge rate. For higher current draw  in the range of 0.5-2~mA, typical batteries such as the CR1220 coin cell can loose as much as 20-40\% of the rated capacity. Finally, periodic high current pulses also reduce the battery capacity. In case of CR1220, a 2.8~mA 2 second pulse even 12 times a day can reduce its capacity by 10\%. The low average current and non-existence of peak currents make backscatter an excellent fit for applications powered by small batteries in comparison to radios. 

\vskip0.05in \noindent{\it ii) Thin film \& printed batteries.} Thin film and printed batteries are attractive due to their small factor but they have significantly lower capacity and peak current capability compared to traditional batteries.  For example, the Molex thin film battery has a peak current rating of 6-8~mA with a capacity of 10~mAh. This peak current can barely support the lowest power mode of a BLE radio and is incompatible with other radio technologies including LoRa, ZigBee and Wi-Fi. Although the peak current requirement can be relaxed by adding large capacitors in parallel to the battery, this adds cost, weight and increase size which counters the value proposition of thin film and printed batteries. Thus, due to the nature of these batteries, backscatter may be a clear winner and in some cases, likely the only choice for communication.

\vskip0.05in \noindent{\it iii) Energy harvesters.} Finally, we consider battery-free solutions which harvest energy from external sources such as RF, solar, vibration and thermal gradient. In addition to limited available power, harvesting has the added limitation of unpredictability in the availability of power. Say, the duty cycle is less than a packet/minute and the average power of a BLE radio is competitive to BLE backscatter solution. Using large batteries, at this duty cycle, backscatter does not provide significant benefits. However, due to the unpredictable nature of energy harvesting, we also need to account for energy per packet requirement for a BLE radio. State of the art Nordic BLE beacons require 33.36~$\mu$J versus 0.2~$\mu$J of energy for Bluetooth backscatter. For an operational voltage of 2~V, this translates to a capacitor size of 83~$\mu$F and 0.5~$\mu$F respectively for a radio and backscatter solution. The time constant for harvester charging the capacitor is directly proportional to capacitor size and backscatter with 100x lower time constant can better withstand variations compared to radios. Imagine someone blocking the RF or solar harvester for a period of time, the backscatter solution would be able to charge the capacitor and operate 100 times quicker than a BLE radio. Additionally, smaller capacitors reduce cost, size and has lower current leakage. 

\section{\normalsize\bf{Open Problems in Backscatter}}
\label{sec:open}
\begin{figure}[t!]

{\footnotesize
	\begin{tabular}{cc}
	\hspace{-0.1in}
	\includegraphics[page = 1, width= 0.375\columnwidth]{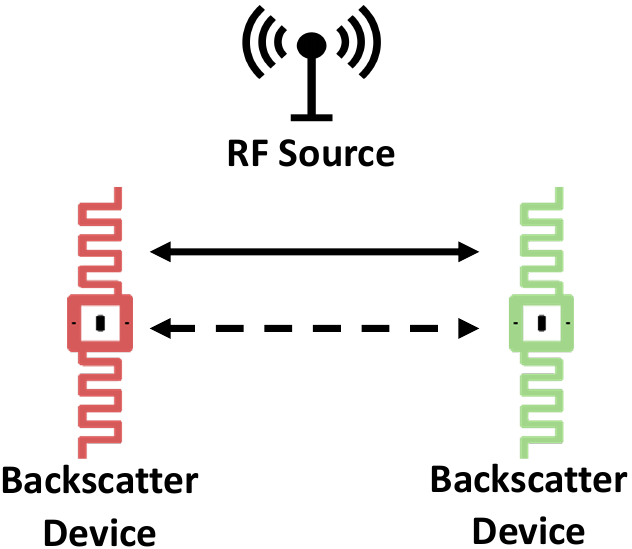} & 
		\hspace{-0.15in}
    \includegraphics[page = 3, width= 0.475\columnwidth]{./figures/paradigm-figures-crop.pdf} \\
	(a) Tag to Tag & (c) Custom Devices \\
	\multicolumn{2}{c}{\includegraphics[page = 2, width= 0.65\columnwidth]{./figures/paradigm-figures-crop.pdf}} \\
	\multicolumn{2}{c}{(b) Commodity Receivers} 
	\end{tabular}
	}
	\vskip -0.15in
	\caption{\textbf{Prominent backscatter architectures.}}
	\label{fig:architectures} 	
	\vskip -0.3in
\end{figure}

Over the past decade, three main backscatter architectures have gained prominence as shown in Fig.~\ref{fig:architectures}. In the first architecture, backscatter  devices communicate with each other by reflecting either ambient signals (e.g., TV~\cite{abc}) or a dedicated carrier source. The challenge with this paradigm is  that receivers on backscatter devices have to be passive to ensure low-power but tend to have poor sensitivity in the range of -40 to -60~dBm which limits operating distances between the backscatter devices. As a result, although ambient backscatter started with tag to tag communication~\cite{abc,abc2014}, researchers moved away from it to receiving backscattered packets on commodity radio receivers (e.g., Wi-Fi~\cite{wifibackscatter}, LoRa~\cite{lorabackscatter}) as shown in Fig.~\ref{fig:architectures}(b). This is attractive since radios have better sensitivity ---  Wi-Fi and Bluetooth have sensitivities of -90 to -100~dBm which is 1000 to 10,000 times lower than a passive receiver~\cite{nsdi16}. Using a LoRa receiver with -130 to -140~dBm sensitivity can further extend the range to 100's of meters~\cite{lorabackscatter}. The final architecture shown in Fig.~\ref{fig:architectures}(c) uses custom signal sources  and custom receivers (potentially full-duplex) to operate the backscatter devices~\cite{ekhonet,batteryFreePhone}. This provides the maximum flexibility but requires changing all devices.

While some of these design (e.g., LoRa backscatter) are mature and deployable, we identify important gaps and open problems in backscatter research.

\vskip0.05in \noindent{\bf 1. Achieving ``truly ambient'' backscatter systems.} 
A truly ambient backscatter system does not require modifications to the wireless infrastructure. The original prototype of ambient backscatter using TV signals~\cite{abc} was true to this paradigm. However, subsequent works on Wi-Fi, Bluetooth and LoRa  introduce additional traffic or require deployment of special devices. Building truly ambient backscatter systems is challenging due to the unpredictability and burstiness of the existing network traffic and the requirement to operate without any  firmware  or hardware modifications to the infrastructure. This is not impossible --- a specific scenario in Wi-Fi backscatter~\cite{wifibackscatter} demonstrated that a truly ambient backscatter system where the tag backscattered ambient Wi-Fi beacons from an unmodified access point and the data was received by a smartphone or any Wi-Fi devices by tracking changes in RSSI measurement, albeit at very low data rate and small range. The key question is: can we  design a system where backscatter tags  transmit data in any indoor environment and the data can be received on a commodity smartphone, without the need to modify existing infrastructure? 

\vskip 0.05in\noindent{\bf 2. Enabling multi-hop backscatter networks.} One of the features that is traditionally associated with sensor networks (e.g., ZigBee) is multi-hop and mesh networking.  Peer to peer communication between two backscatter devices was used in the original ambient backscatter work that uses ambient TV signals~\cite{abc,abc2014}. However, the system operated at short distances and low data rates and required to be close to the TV station and do not work indoors. Achieving multi-hop backscatter systems is challenging and many of the subsequent works  have shied away from addressing this problem. Recent work~\cite{barnet} have taken a difference approach of introducing a dedicated UHF carrier source  but is limited to two-hops, achieving only 200 bps with tag-to-tag distances of 40~cm~\cite{xtandem}. It is unclear if at these distances it would benefit the network throughput in comparison to directly transmitting to the access point.   Designing multi-hop backscatter system that can achieve information theoretically provable throughput gains is an open problem.

\vskip 0.05in\noindent{\bf 3. Connectivity for  computational materials.} Communication innovations has been limited to engineers who are experts in electronics and radios. To truly democratize the vision of  ubiquitous connectivity we need to be able to integrate communication capabilities with everyday objects and materials (e.g., plastics) without the need for electronics or batteries. Recent work on printed Wi-Fi shows the one can enable 3D printed wireless sensors and input widgets using plastic materials by leveraging backscatter~\cite{printedwifi}. Backscatter is achieved by using mechanical motion to change the properties of the antenna, which results in very bulky objects since mechanical actuators do not  scale down as transistors. There is significant research to be done on using meta-materials with interesting mechanical properties to achieve small computational materials that can communicate using backscatter.

\vskip 0.05in\noindent{\bf 4. Programmable smart dust.} While there has been interest in creating small weight, size and power smart dust sensors, the need for spinning out tiny custom radio ICs has limited smart dust from taking off beyond a few select IC researchers. One of the advantages of using backscatter is that it does not require a large number of components (e.g., external oscillators) and hence can deliver connectivity using only a microcontroller.  Further, advancements in technology scaling, dynamic voltage scaling, sub-threshold operation and power gating techniques have drastically reduced power consumption and size of micro-controllers. As an example, the MSP430, one of the lowest power MCU in 2013 consumed 100~$\mu$A/MHz compared to 6~$\mu$A/MHz current consumption of the latest Apollo 3 by Ambiq. Further some of the commercial microcontrollers  weight as low as a few milligrams.  Finally, due to recent advances in semi-conductor processes and packaging technologies, power and size of RF switches have also reduced. This enables light-weight and low-power microcontroller-based sensor systems that weigh less than 100 mg~\cite{livingiot}. We believe that pursuit of power  optimized microcontroller-based backscatter designs is a worthwhile direction. While existing efforts like WISP are focused on low rates, a similar effort should be undertaken to build microcontroller-based designs for higher data rate systems.

\vskip 0.05in\noindent{\bf 5. Millimeter, Terahertz and Quantum backscatter.} A key advantage of backscatter communication is that it does not require generating the carrier frequency. In general, the energy requirements to generate the carrier frequency increase  with operating frequency. Thus the energy consumption of generating millimeter and terahertz signals is non-negligible. An under-explore research direction is to create backscatter network at the millimeter and terahertz frequencies. An advantage of these frequencies is that owing to their small wavelength, we can pack in more antennas in a small area and hence can achieve very directional beam. This is beneficial since the signal source can beam-form the signal in the direction of the backscatter device and achieve longer ranges than at  2.4~GHz. Further, creating a multi-antenna backscatter device can beam-form towards the receiver and can also multi-cast to different receivers. While, there has been some initial work~\cite{freidl2017mm, kimionis2017millimeter}, better exploring  backscatter networks at these high frequencies is a high reward research direction. Another interesting direction is to achieve secure communication using quantum backscatter.

\vskip 0.05in\noindent{\bf 6. High-resolution audio and video streaming.} Streaming applications such as audio and video streaming which require continuous transmission of data are a great fit for backscatter as backscatter shines over radios when the duty cycle of transmission is significantly higher. While analog backscatter approaches which eliminate digital operations have enabled battery-free phones~\cite{batteryFreePhone}, the resulting audio quality is significantly degraded. Similarly recent work achieves low-resolution video streaming at a low-frame rate using backscatter~\cite{nsdi2018camera}. An important research question is to design high-resolution audio and video streaming using backscatter at orders of magnitude lower power than existing radio based systems. This could require an inter-disciplinary approach across IC design as well as video and audio compression to design the appropriate low-power compression algorithms that minimize the bandwidth usage. Another approach is to leverage advances in video super resolution at the access point.

\vskip 0.05in\noindent{\bf 7.  Acoustic backscatter on surfaces and in air.} An unexplored research direction is to backscatter acoustic signals on surfaces and in air. Acoustic backscatter  has traditionally been limited to communication through tissue for implants and underwater applications. However, surface transducers provide an efficient mechanism to communicate over the surface of the human body or other material (e.g., metal). Using backscatter on signals from such surface transducers can enable low-power surface communication systems. More importantly, acoustic backscatter could be beneficial if demonstrated in air. Smart speakers and phones are equipped with speakers and microphones that can transmit and receive acoustic signals. This can enable low-power devices that can use backscatter to communicate with existing smart devices. This is beneficial since backscattering to a single device requires full-duplex capabilities that are challenging with radio systems. However, speakers and microphones on commodity devices can achieve full duplex capabilities easily.

\vskip 0.05in\noindent{\bf 8. Ultra-wide band and frequency-agnostic backscatter.}  We have seen a significant change in the mobile phone landscape in the last few years with the introduction of  ultra-wideband (UWB) radios in Apple iPhones. This has been offered as a solution to provide ``spatial awareness'' since UWB radios are known to provide  accurate distance and angle information. Exploring UWB backscatter systems that, is a promising research direction, in particular if this can be achieved while being compatible with the Apple UWB protocols. This can enable low-power tags that can be attached to objects and integrate with augmented reality (AR) applications while communicating  with mobile devices that support UWB protocols. Another research direction is in creating  a frequency agnostic backscatter system capable of operating across different protocols and frequency bands from TV, cellular, 2.4 GHz, 5.8 GHz to millimeter  bands and has the potential to operate universally across locations, countries and applications. \cite{alanson} has made preliminary progress in this direction by using a wide-band antenna on the backscatter device. However, this system was limited in data rate and  a maximum frequency of 900 MHz. Achieving a truly frequency-agnostic system require innovations in hardware as well as low-power algorithms to dynamically identify which frequency bands have the strongest signal at low power.

\vskip 0.05in\noindent{\bf 9. Improving RF harvesting and the downlink.} While backscatter can be used with batteries,  RF harvesting is still a popular choice for building battery-free systems since we can use the same antenna for both power and communication. However, the sensitivity --- minimum RF signal strength --- required for RF harvesting to work is orders of magnitude worse than backscatter communication. Can we improve harvester sensitivity  from -20-30~dBm to -40-50~dBm  by a combination of leakage power reduction, technology scaling and sub-threshold operating using voltage scaling?  This would be notable as every 6~dB improvement in sensitivity doubles the operating distance.  Finally,  a downlink from an access point to the backscatter device is needed to schedule transmissions by endpoints, mediate the wireless channel, send packet acknowledgement, implement rate adaptation and schedule packet re-transmissions in case of packet errors. Since, active receivers consume orders of magnitude higher power, passive receivers based on envelope detectors which decode on-off keying modulation while consuming micro-watts of power have been exclusively used. This however are not resilient to  interference and are not supported by commodity mobile devices.  Prior work conveys information in the presence and absence of WiFi packets~\cite{wifibackscatter},  modulating the packet length  packet~\cite{wifibackscatter, hitchhike} or reverse engineering OFDM~\cite{interscatter} but with limited data rates. Is it possible to  have a downlink that is as good (in terms of rate and energy efficiency) as the backscatter uplink? Is it possible to even consider passive-like receivers that can decode legacy protocols like Wi-Fi/Bluetooth?

\vskip 0.05in\noindent{\bf 10. Backscatter network and MAC  protocols.} Backscatter research is still in its infancy with primary focus on novel physical layer design. Recent research has proposed  to solve some of the pressing networking and MAC-layer challenges of backscatter~\cite{netscatter}. However, the majority of implementation and evaluation has been focused on the physical layer with piecemeal evaluation. We hope that with advancements, the physical layer is robust enough to start focusing on some of the critical network and MAC layer challenges of backscatter devices: provisioning devices in the network, perform carrier sense, network management, polling of devices and many more. It is important to have a large scale backscatter system that is available to researchers across the world, perhaps similar to Planetlab, being evaluated and deployed where networking questions are answered with robust evaluation and comparison across protocols. 
\section{\normalsize\bf{Conclusion}}
To conclude, this paper first makes the case for decoupling backscatter communication from RF power harvesting and provides a systematic design space for backscatter applications. We then provide  ten important open research problems in backscatter networking that we believe are essential to deliver on the promise of low-power wireless connectivity for the next billion devices. 

\balance
\bibliographystyle{abbrv} 
\begin{small}
\bibliography{hotnets2020}
\end{small}

\end{document}